\documentclass[prc,twocolumn,superscriptaddress,showpacs,twoside,floatfix]
{revtex4-1}

\usepackage{amssymb,epsfig}
\usepackage{color}
\begin{document}

\title{Exploring manifestation and nature of a dineutron in two-neutron
emission using a dynamical dineutron model}

\author{L.V.~Grigorenko}
\affiliation{Flerov Laboratory of Nuclear Reactions, JINR,  RU-141980 Dubna,
Russia}
\affiliation{National Research Nuclear University ``MEPhI'', Kashirskoye shosse
31, RU-115409 Moscow, Russia}
\affiliation{National Research Centre ``Kurchatov Institute'', Kurchatov sq.\ 1,
RU-123182 Moscow, Russia}
\author{J.S.~Vaagen}
\affiliation{Institute of Physics and Technology, University of Bergen, N-5007 Bergen, Norway}
\author{M.V.~Zhukov}
\affiliation{Department of Physics, Chalmers University of Technology, S-41296 G\"{o}teborg, Sweden}

\begin{abstract}
Emission of two neutrons or two protons in reactions and decays is often discussed in terms of ``dineutron'' or ``diproton'' emission. The discussion often leans intuitively on something described by Migdal-Watson approximation. In this work we propose a way to formalize situations of dineutron emission. It is demonstrated that properly formally defined dineutron emission may reveal properties which are drastically  different from those traditionally expected, and properties which are actually observed in three-body decays.
\end{abstract}

\maketitle


\section{Introduction}


The idea of final state interaction (FSI) treatment in Migdal-Watson approximation \cite{Watson:1952,Migdal:1955} is one of the basic concepts of nuclear reaction theory. In this approach the low-energy modification is predicted in the relative energy spectra of decay fragments interacting in the final state. This modification is related to the spectrum of fragments which defines pragmatic use of the approach. The low-energy cross section in the corresponding channel (with energy $E$) is factorized as
\begin{equation}
\frac{d \sigma}{dE} \sim F_{FSI}(E) \, F_{PV}(E_T,E) \, ,
\label{eq:sigma-mw}
\end{equation}
where $F_{PV}$ is the ``phase volume'' contribution. In three-body decays with total decay energy $E_T$ this term is
\begin{equation}
F_{PV}(E_T,E) = \sqrt{E(E_T-E)} \, .
\label{eq:sigma-pv}
\end{equation}
The FSI term is obtained as
\begin{equation}
F_{FSI}(E) = \frac{1}{C_l^2} \frac{1}{2ME \,[ \cot ^2 \delta_l(E)+1]} \, ,
\label{eq:f-pv}
\end{equation}
where $M$ is the reduced mass in the channel of interest. The Coulomb penetration factor $C_0$ for $l=0$ is defined via Sommerfeld parameter $\eta$
\begin{equation}
C_0^2(E) =  \frac{2\pi\eta(E)}{\exp[2 \pi \eta(E)]-1} \,, \quad
\eta(E) = \frac{Z_1Z_2 \alpha}{\sqrt{2E/M}}\, ,
\label{eq:coul-penetr}
\end{equation}
and tends to unity in the case of neutral particles. For $s$-wave interaction of neutral particles the FSI term can be approximated in terms of the effective range approach as
\begin{equation}
F_{FSI}(E) = \frac{a^2}{1+2ME \, a^2} \, ,
\label{eq:f-fsi}
\end{equation}
where $a$ is an $s$-wave scattering length. Thus in the original Migdal-Watson approximation the low-energy modification of the cross section  is sensitive to just one parameter: the scattering length in the channel of interest. For emission of two neutrons such a behavior of relative energy distribution gives rise to the notion of a ``dineutron'' particle as a specific object of research.

In spite of the fact that the ``dineutron'' idea is quite old there remain several aspects of theoretical importance, explored in current studies.

\noindent (a) $d(n,np)n$ reaction (and analogous reactions) as a tool to study $n$-$n$ scattering length. There exists a problem of charge symmetry breaking for $n$-$n$ and $p$-$p$ channels (difference in the $s$-wave scattering lengths). Since it is very difficult to study neutron-neutron collisions directly, indirect methods have to be applied (e.g.\ Refs.\ \cite{Gardestig:2009,Witala:2010,Couture:2012} and Refs.\ therein).

\noindent (b) A Hanbury-Brown-Twiss ``HBT interferometry''-like approach for high-energy collisions \cite{Sinyukov:1998,Lisa:2005}. This ``femtoscopy'' approach allows to extract characteristics of the collision region, from which the emission of correlated particles is observed. It was suggested in Refs.\ \cite{Marques:2000,Marques:2001,Marques:2002} that an analogous ``HBT interferometry''-like approach can be used for reactions with light exotic nuclei to extract the radial characteristics of a neutron halo. In this work we try to find out which information can actually be extracted in such studies.

\noindent (c) ``Dineutron emission'' in decays of light exotic nuclei. The even-neutron systems beyond the neutron dripline typically decay via direct emission of two neutrons. This process is sometimes discussed in terms of ``dineutron emission''. The declared discovery of ``dineutron emission'' in decay of $^{16}$Be has recently produced a heated discussion, see Refs.\ \cite{Spyrou:2012,Marques:2012,Spyrou:2012a}. In this work we try to clarify this discussion by improved assessment of its theoretical constituents.

Some sources of current confusion in the discussion of a ``dineutron'' are as follows.

\noindent (i) It is important to note that we consider emission of a ``dineutron'' with \emph{low} total decay energy $E_T$. However, it could have two physically very different sources: two-neutron \emph{decays} of low-lying resonant states or \emph{reactions} leading to population of low-lying three-body continuum. The formal description of these situations is very different.

\noindent (ii) The ``dineutron'' is often described as a spatial correlation of two neutrons in the nuclear interior caused by the pairing interaction. It is often erroneously assumed that such a compact spatial configuration should exhibit itself as low-energy enhancement in the spectrum of two neutrons. This vision contradicts the uncertainty principle: a short-distance correlation should correspond to large relative momenta. So, the considerable large-momentum enhancement by spatial ``dineutron'' (caused by pairing) should be effectively overcome in the process of decay by the low-momentum enhancement (caused by the final state interaction). It appears that the issue of such an interplay defines applicability of the Migdal-Watson approximation and it is especially addressed in this work.

As an illustrating case of dineutron emission we have selected the ground state decay of $^{26}$O. The latter has recently attracted considerable attention, both experimental  \cite{Lunderberg:2012,Caesar:2013,Kohley:2015,Kondo:2016} and theoretical \cite{Volya:2006,Grigorenko:2013,Volya:2014,Hagino:2014,Hagino:2014b,Grigorenko:2015b,Fossez:2017}. In our previous works the two-neutron emission from the $^{26}$O g.s.\ has been studied in various theoretical approximations including sophisticated three-body decay and reaction models \cite{Grigorenko:2011,Grigorenko:2013,Grigorenko:2015b}. Generally, we find the dineutron approximation too poor and that complete three-body calculations (treating all pairwise final state interactions in the system on the same ground) are required to deal with three-body decays in all their complexity. However, using the limited model we explore two important tasks.

\noindent (1) We attempt to clarify the question how the dineutron emission should look like if such a process takes place in reality for whatever reason. The results of these studies could be very discouraging for those who utilize this concept without sufficient theoretical background.

\noindent (2) We solve some methodological problems of our approach to three-body decays in fully controllable conditions. This helps to further validate our results concerning several complicated aspects of true $2n$  emission.

It should be noted that effects of nucleon-nucleon interactions on three-body $2p$ or $2n$ decays were investigated theoretically in several recent works \cite{Grigorenko:2009c,Grigorenko:2013,Oishi:2017}. These works demonstrated important effects of this aspect of final state interaction on the decay widths and correlations. In this work we study in a sense an opposite problem: starting from nucleon-nucleon FSI (given by default as the only long-range effect) we try to understand which kind of information about nuclear interior can ``pass'' through such a ``filter''.


\section{Theoretical approximations}


The discussion of ``dineutron emission'' is often lacking clarity because the object is loosely defined. For three models described in this Section it is defined which physical situation is considered in each case. The first two models described below are commonly used (or implied to be used), while the third model is developed in this work.


\subsection{Trivial two-body dineutron emission}
\label{sec:2n-triv}


\begin{figure}
\begin{center}
\includegraphics[width=0.47\textwidth]{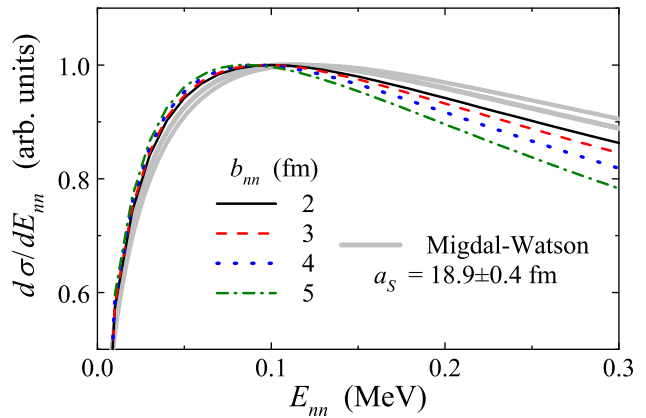}
\end{center}
\caption{Neutro-neutron energy correlation for dineutron emission by a two-body source function. All curves are normalized to unity maximum value.}
\label{fig:nn-trivial}
\end{figure}

Let us consider dineutron emission from a static source $\Phi$ with rms radius $b_{nn}$. The decay probability can in this case be defined via the outgoing flux
\begin{equation}
\frac{d \sigma}{d E_{nn}} \sim j_{E_{nn}} = \frac{1}{M} \left. \text{Im} \left[ \Psi^{(+)\dagger}_{E_{nn}}(\mathbf{r}) \nabla \Psi^{(+)}_{E_{nn}}(\mathbf{r}) \right] \right|_{r \rightarrow \infty} \,,
\label{eq:ds-denn}
\end{equation}
associated with wave function (WF)
\begin{equation}
\Psi^{(+)}_{E_{nn}}(\mathbf{r}) = \int d^3 r' \, G^{(+)}_{E_{nn}}(\mathbf{r};\mathbf{r}') \, \Phi(\mathbf{r}')\,.
\label{eq:nn-trivial}
\end{equation}
For the source function $\Phi$ defined by first oscillator WF $\phi_{00}$
\begin{eqnarray}
\Phi(\mathbf{r}) = \sum_{lm} \frac{\phi_{nl}(r)}{r}  \, Y_{lm}(\hat{r})
\,, \nonumber \\
\phi_{00}(r)  =  \frac{1}{b_{nn}^{3/2}} \left(\frac{54}{\pi}\right)^{1/4} \, r \, \exp \left(-\frac{3r^2}{4b_{nn}^2} \right)\,,
\label{eq:source-pp}
\end{eqnarray}
one gets the nucleon-nucleon low-energy correlations shown in Fig.\ \ref{fig:nn-trivial}. The radius parameter $b_{nn}$ in (\ref{eq:source-pp}) is defined in such a way that it is equal to the source rms radius.

A simple nucleon-nucleon interaction is used in this work, acting only in the $s$-wave of two neutrons and defined by a potential with Gaussian formfactor
\begin{equation}
V_{nn}(r) = V^{(0)}_{nn} \exp[-(r/r^{(0)})^2]\,.
\label{eq:pot-nn}
\end{equation}
For depth $V^{(0)}_{nn}=-31$ MeV and width $r^{(0)} =1.8$ fm this potential produces the scattering length $a=-18.7$ fm. The Migdal-Watson expression provides the peak in the $n$-$n$ energy correlation spectrum at about 115-120 keV (depending on the scattering length experimental uncertainty $a_S=18.9\pm 0.4$ fm). The peak produced for emission off a static source is somewhat different: it is located at somewhat lower energies of 70--100 keV for realistic ``sizes'' of the dineutron correlation of $b_{nn}=3-7$ fm. Also the shapes of the spectrum are quite sensitive to the radius parameter $b_{nn}$. Look for further discussion of this issue in Section \ref{sec:mw-valid}.

This approximation provides some qualitative idea about what could be called ``dineutron emission''. However, there exists a unique situation in which the description of two-neutron emission by Eq.\ (\ref{eq:nn-trivial}) becomes adequate, see the end in the next Subsection.


\subsection{Static three-body dineutron model}
\label{sec:2n-stat}


Next consider the dineutron emission from a static source $\Phi$ consisting of two nucleon WFs occupying some orbital configurations:
\begin{equation}
\Phi_{JM}(\mathbf{r}_1,\mathbf{r}_2) = [\Phi(\mathbf{r}_1) \otimes \Phi(\mathbf{r}_2) ]_{JM} \,.
\label{eq:souce-3b}
\end{equation}
This model we refer to in the following as a static dineutron model (S2nM). After conversion of this source to Jacobi coordinates (so-called ``T'' system)
\begin{equation}
\mathbf{X} = \mathbf{r}_1 - \mathbf{r}_2 \,,\quad \mathbf{Y} = \frac{A}{2(A-2)}\,(\mathbf{r}_1 + \mathbf{r}_2) \,,
\label{eq:jac-t}
\end{equation}
($A$ is the mass number of the system of interest) the dineutron emission from this source can be treated exactly
\begin{equation}
\Psi^{(+)}_{E_T} = \frac{1}{\hat{T}_3+V_{nn}-E_T+i\epsilon} \, \Phi \,,
\label{eq:nn-sdm1}
\end{equation}
\begin{eqnarray}
\Psi^{(+)}_{E_T,JM}(\mathbf{X},\mathbf{Y}) =  \int d^3 X' d^3 Y'\, G^{(+)}_{E_T}(\mathbf{X},\mathbf{Y};\mathbf{X}',\mathbf{Y}') \nonumber \\
\times \, \Phi_{JM}(\mathbf{X}',\mathbf{Y}')\,.\quad
\label{eq:nn-sdm2}
\end{eqnarray}
Here the three-body kinetic energy $\hat{T}_3$ is given by
\[
\hat{T}_3 = \frac{\hat{P}_x^2}{2M_x} + \frac{\hat{P}_y^2}{2M_y}\, , \;\;\;
M_x = \frac{M_n}{2} \, , \;\;\; M_y = \frac{2(A-2)}{A}\,M_n\,,
\]
where $M_n$ is nucleon mass and $\hat{P}_x$, $\hat{P}_y$ are momentum operators conjugated to Jacobi coordinates $X$, $Y$ (\ref{eq:jac-t}). The above three-body Green's function $G^{(+)}_{E_T}$ can be given in a simple analytic form
\begin{eqnarray}
G^{(+)}_{E_T}(\mathbf{X},\mathbf{Y};\mathbf{X}',\mathbf{Y}') = \frac{1}{2 \pi i} \int dE_x \, G^{(+)}_{E_x}(\mathbf{X};\mathbf{X}') \nonumber \\
\times \, G^{(+)}_{E_T-E_x}(\mathbf{Y};\mathbf{Y}')\,, \quad
\label{eq:gf-simpl}
\end{eqnarray}
where the $Y$ variable Green's function corresponds to plane wave propagation, while the $X$ variable Green's function incorporates the $n$-$n$ final state interaction.

The model is called ``static'' in the sense that the properties of the source are totally decoupled from the properties of the final state interaction. The realistic scenario for such a model is sudden  removal of the core from a two-nucleon halo system. This is not an improbable scenario for high-energy direct knockout reactions. For example, it was demonstrated in Ref.\ \cite{Sidorchuk:2010} that about $50 \%$ of the $^{4}$He($^{6}$He,$2 \alpha$) cross section, even at not very high beam energy of $\sim 25$ AMeV, can be related to quasi-free knockout of the $\alpha$ core from the $^{6}$He nucleus. In such a case the source function $\Phi_{JM}$ can be immediately related to the WF of the valence halo nucleons, paving way for studies of this WF structure.

It should be noted that there exists only one approximation in which the S2nM is reduced to the ``trivial dineutron emission'' of the previous Section. This is realized if the source function can be written in the factorized form:
\begin{equation}
\Phi(\mathbf{r}_1,\mathbf{r}_2) \equiv \Phi(\mathbf{X}) \, \Phi(\mathbf{Y})\,.
\label{eq:factor}
\end{equation}
The one and only case when this is possible, is when the $J=0$ source is represented by two lowest $s$-wave oscillator WFs $\phi_{nl}(\mathbf{r})$
\begin{equation}
\Phi(\mathbf{r}_1,\mathbf{r}_2) \equiv \phi_{00}(\mathbf{r}_1) \, \phi_{00}(\mathbf{r}_2)\,.
\label{eq:factor-2}
\end{equation}
In this case all the information contained in the nucleon-nucleon momentum distribution is fully described by Eq.\ (\ref{eq:nn-trivial}). This is exactly the situation considered in the applications of the HBT interferometry ideas to high-energy reactions: Emission of independent particles from thermal source with Gaussian radial formfactor is formalized exactly by this model. For sources stemming from low-energy nuclear reactions this approximation is too poor because of variety of radial nucleon WFs deviating from Gaussian shapes and variety of angular momentum couplings defined by the investigated valence nucleon configurations.


\subsection{Dynamic three-body dineutron model}
\label{sec:2n-dynam}


In the case of resonance state decays the S2nM can not be a reasonable approximation being associated with a certain reaction class. For resonant states in the limit of infinite lifetime the emission process should become totally insensitive to the population mechanism (as we have mentioned above the S2nM can be associated with a certain reaction class). As an adequate dynamic approximation to the dineutron emission we now consider the following Dynamic Dineutron Model (D2nM). The decay of a three-body system is considered by solving a Schr\"odinger equation for WF $\Psi^{(+)}$ with purely outgoing wave boundary conditions and complex energy
\begin{equation}
(\hat{H}_3-E_T+i \Gamma/2)\Psi^{(+)}_{E_T} = 0 \;, \quad \hat{H_3} = \hat{T}_3 + \hat{V}_3(\rho) + V_{nn}(X) \,.
\label{eq:shred-ddm}
\end{equation}
The three-body Hamiltonian $\hat{H_3}$ contains nucleon-nucleon potential $V_{nn}$, kinetic energy term $\hat{T}_3$, and phenomenological three-body potential $\hat{V}_3$. The latter has short-range behavior in the hyperradius $\rho$, which should guarantee abscence of other long-range effects than those connected with $V_{nn}$.

To solve the three-body Schr\"odinger equation Eq.\ (\ref{eq:shred-ddm}) we use the Hyperspherical Harmonics (HH) method and the iterative procedure developed in Ref.\ \cite{Grigorenko:2007}. In the first step we use the hyperspherical harmonics method with ``box'' outgoing boundary conditions also defining the real part of the decay energy $E_T$:
\[
(\hat{H}_3-E_T)\Psi_{\text{box}} = 0 \,.
\]
Then the WF with outgoing asymptotic is derived solving the inhomogeneous equation
\[
(\hat{H}_3-E_T)\Psi^{(+)}_{E_T} = - \, (i \Gamma/2) \, \Psi_{\text{box}}
\]
The obtained solution $\Psi^{(+)}_{E_T}$ may have problems with convergence, connected with effective ``long-range'' character of nucleon-nucleon interaction in $s$-wave. Near perfect work-around for such problems exists for simplified Hamiltonians, which include only one or two final state interactions and therefore there exists an analytic Green's function. In brief, we can rearrange Eq.\ (\ref{eq:shred-ddm}) in the following way:
\begin{equation}
\Psi^{(+)}_{E_T} = - \, \frac{1}{\hat{T}_3+V_{nn}-E_T+i\Gamma/2} \, \hat{V}_3(\rho) \, \Psi^{(+)}_{E_T} \,.
\label{eq:shred-split}
\end{equation}
In the limit $\Gamma \ll E_T$ we again get in the right-hand side of Eq.\ (\ref{eq:shred-split}) the analytically known Green's function $\hat{G}^{(+)}_{E_T}$ of Eq.\ (\ref{eq:gf-simpl}), which makes possible iterative improvement of the solution  $\Psi^{(+)}_{E_T}$ providing the ``corrected'' WF $\Psi^{(+)}_{E_T,\text{corr}}$
\begin{equation}
\Psi^{(+)}_{E_T,\text{corr}} = - \, \hat{G}^{(+)}_{E_T} \, \hat{V}_3(\rho) \,\Psi^{(+)}_{E_T}\,.
\label{eq:wf-corr}
\end{equation}
Convergence of the procedure is guaranteed for the short-range potential $\hat{V}_3(\rho)$. There is also a simple criterion to check the consistency of the procedure: The resonant state widths and three-body momentum distributions obtained before and after some number of iterations of the ``correction'' step should coincide.


\section{What affects dineutron structure?}
\label{sec:struct}



\subsection{Structure effects in D2nM}
\label{sec:struct}


In this Section we try to isolate the internal nuclear structure effects on the dineutron emission. In D2nM we form the required structure by selection of the three-body potential $\hat{V}_{3}(\rho)$ to be different for hyperspherical components with different $K$ values
\begin{equation}
\hat{V}_{3}(\rho) = \sum_{K} \frac{V_{3,K}}{1+\exp[(\rho-a_{\rho})/d_{\rho}]} \; \hat{P}_K\,.
\label{eq:v3}
\end{equation}
Here a Woods-Saxon formfactor is chosen, while $\hat{P}_K$ is projector on the states with definite $K$ values.

We consider primarily the lowest excitations with $J^{\pi}=0^+$. In the proposed model the lowest energy three-body $0^+$ WF has only one component with $L=0$, $S=0$, $l_x=0$, $l_y=0$, which corresponds to a dineutron in $s$-wave  motion relative the core. The potential parameters used in the calculations are listed in Table \ref{tab:potent}. The total decay energy $E_T$ for each calculation is controlled just by the one running parameter $V_{3}$.

\begin{table}[b]
\caption{Depth parameters of the three-body potential $\hat{V}_{3}(\rho)$ in Eq.\ (\ref{eq:v3}) which are used for calculations providing different dominant $[l^2]_0$ configurations. Geometry parameters $a_{\rho}=4$ fm and $d_{\rho}=0.8$ fm were also used.}
\begin{ruledtabular}
\begin{tabular}[c]{cccc}
 Case     &  $V_{3,0}$ & $V_{3,2}$ & $V_{3,4}$   \\
\hline
 $[s^2]$  &   $V_3$    & 0         & 0        \\
 $[p^2]$  &  200       &   $V_3$   & 0    \\
 $[d^2]$  &  200       & 200       &  $V_3$  \\
\end{tabular}
\end{ruledtabular}
\label{tab:potent}
\end{table}

\begin{figure*}
\begin{center}
\includegraphics[width=0.97\textwidth]{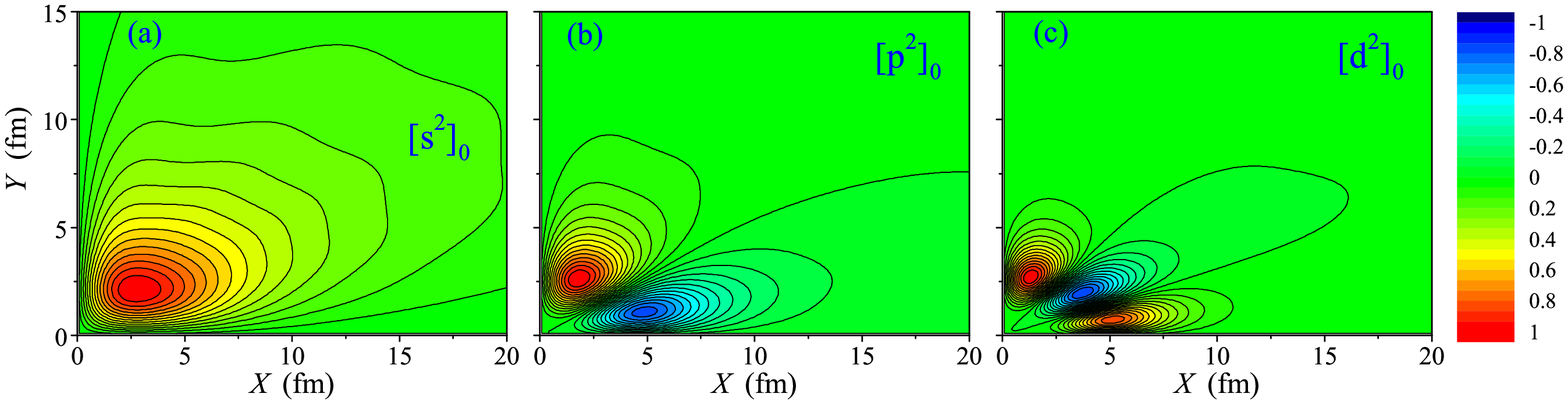}
\includegraphics[width=0.343\textwidth]{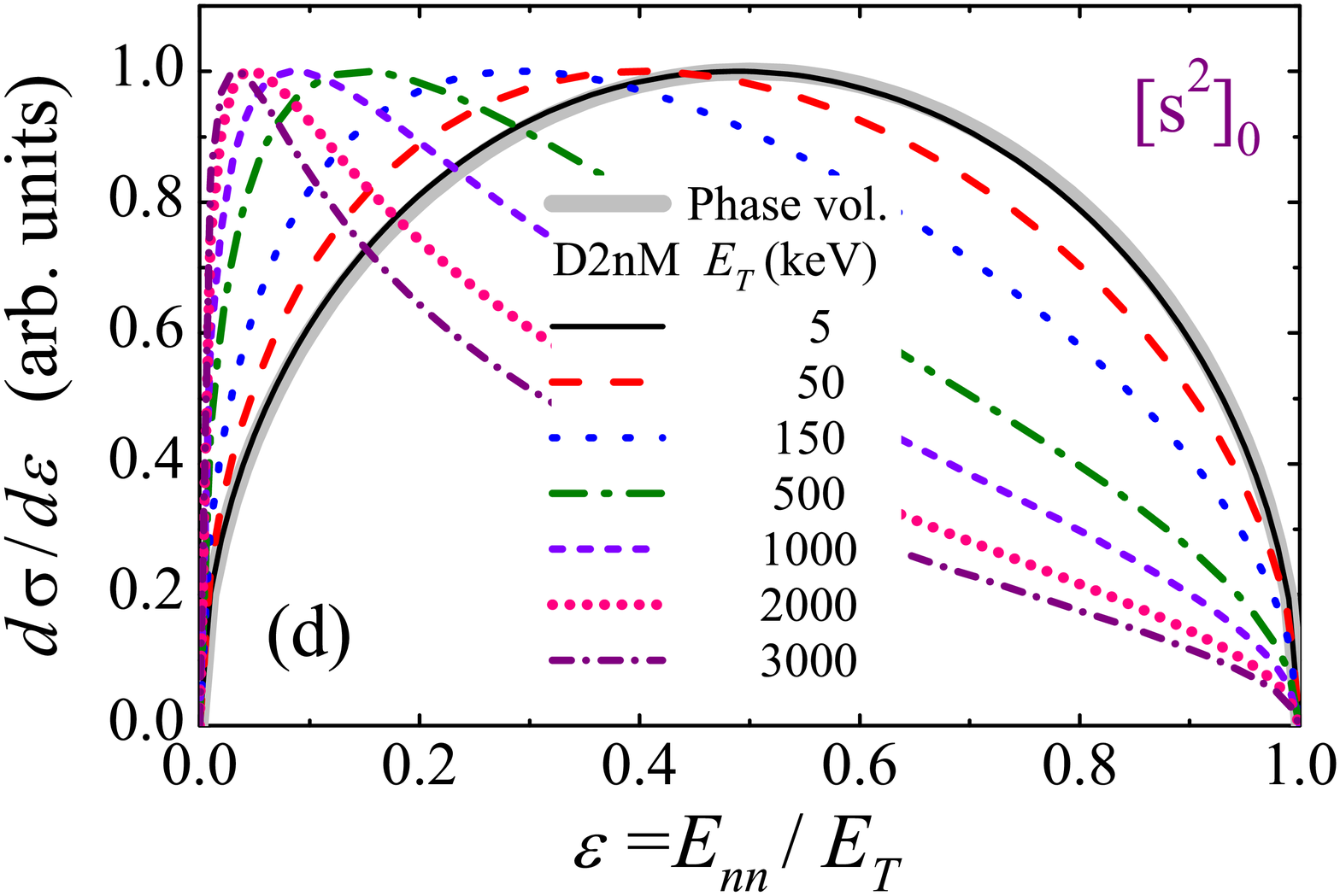}
\includegraphics[width=0.31\textwidth]{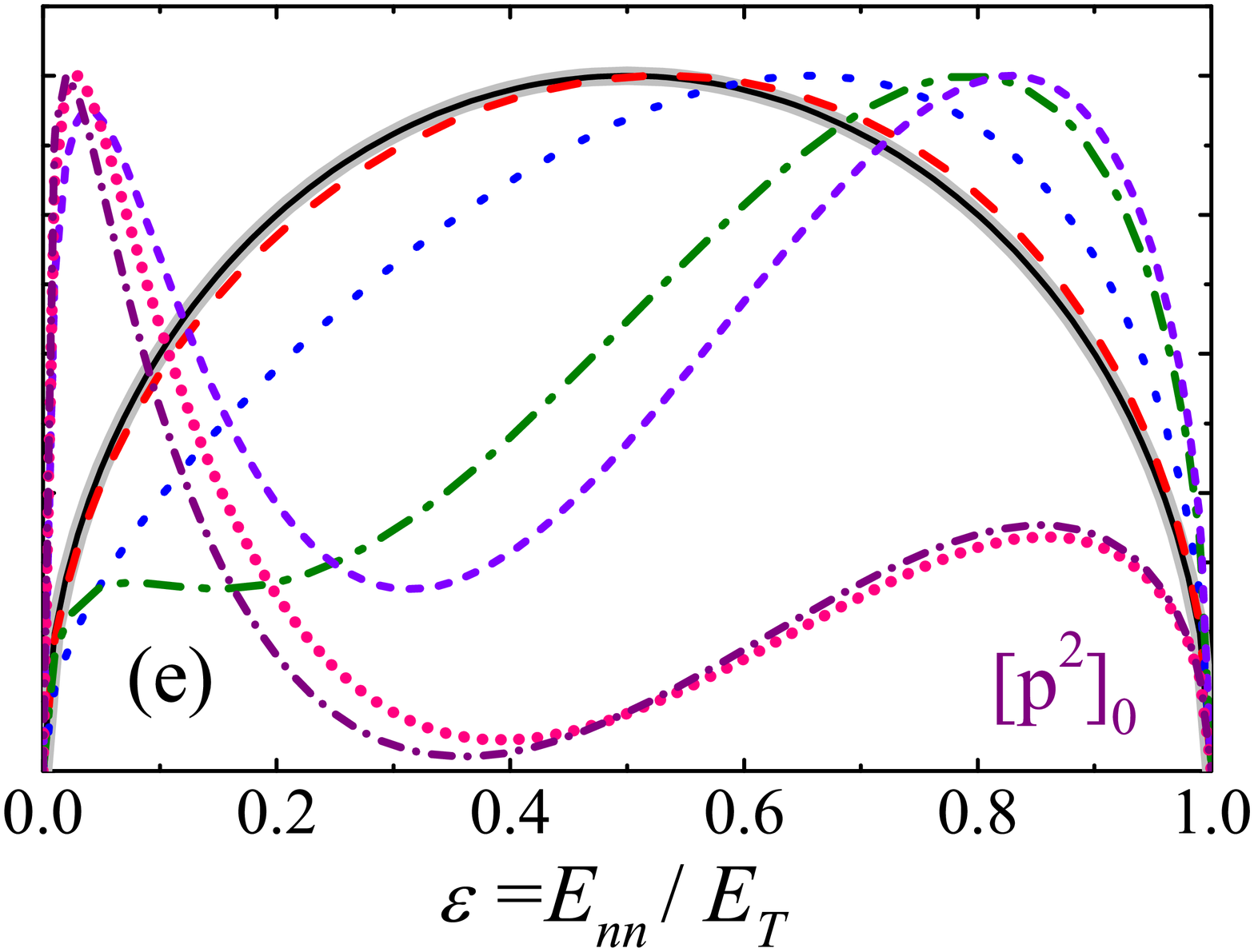}
\includegraphics[width=0.31\textwidth]{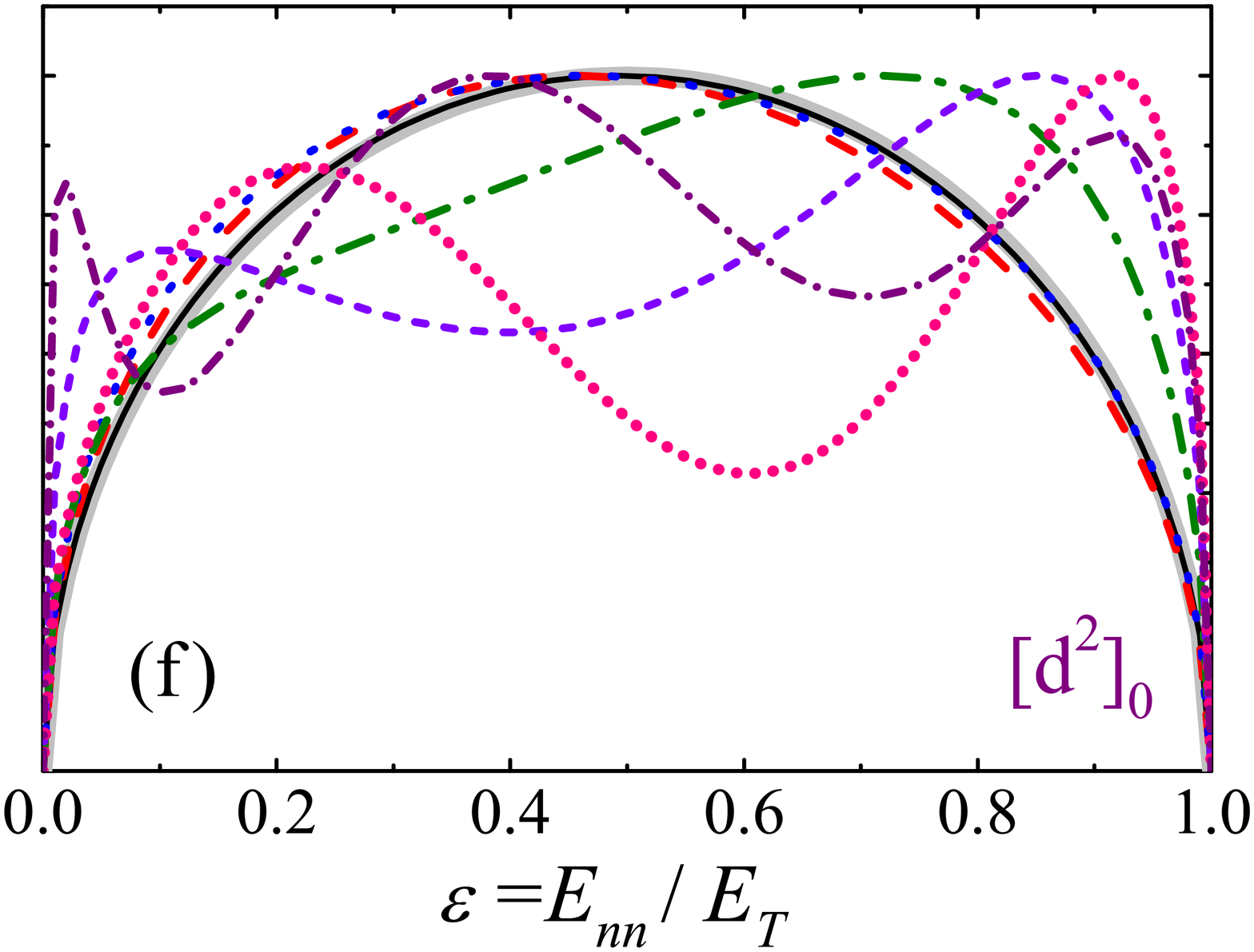}
\end{center}
\caption{Spatial correlations in the internal region are illustrated by the real part of the decay WF $\Psi^{(+)}_{E_T}$ for $E_T=50$ keV. Panels (a), (b), and (c) correspond to calculations showing, in the internal region, dominant $[s^2]_0$, $[p^2]_0$, and $[d^2]_0$ configurations, respectively. Energy correlations between two neutrons for different total decay energies $E_T$ are given in corresponding panels (d), (e), and (f). Gray lines show the three-body phase volume distribution. All surfaces and curves are normalized to unity maximum value.}
\label{fig:sructure-eff}
\end{figure*}

The results are shown in Fig.\ \ref{fig:sructure-eff}. The upper panels illustrate the spatial correlations in the ``T'' Jacobi system ($X$ is distance between two neutrons and $Y$ is distance between $n$-$n$ center of mass and heavy fragment). The selection of a structure strongly dominated by the $K=0$, $K=2$, or $K=4$ component by potential in Table \ref{tab:potent}, leads to corresponding population of very pure $[s^2]$, $[p^2]$, and $[d^2]$ quantum configurations. Domination of these structures is clearly seen in Fig.\ \ref{fig:sructure-eff} (a), (b), and (c) as presence of one, two, or three peaks of the WF in the internal region. Such correlation patterns are connected with Pauli principle and are often referred to as ``Pauli focusing'', Ref.\ \cite{Danilin:1988}.

The energy correlations between two emitted neutrons, for WFs with corresponding internal structures, are illustrated in lower panels of Fig.\ \ref{fig:sructure-eff}. They are expressed in terms of fractional energy variable
\[
\varepsilon=E_{nn}/E_T \, .
\]
It is shown that for decay energies $E_T<150$ keV the obtained correlations are relatively close to the three-body ``phase volume''
\[
d \sigma / d \varepsilon \sim \sqrt{\varepsilon (1-\varepsilon)} \,.
\]
Thus, for such decay energies the $n$-$n$ FSI is not strong enough to noticeably modify the phase volume distribution. Only at about $E_T \sim 500$ keV do the correlation patterns begin to deviate considerably from the phase volume decay. At this and higher energies the structure effects are seen to play a dominant role.

\begin{figure}
\begin{center}
\includegraphics[width=0.47\textwidth]{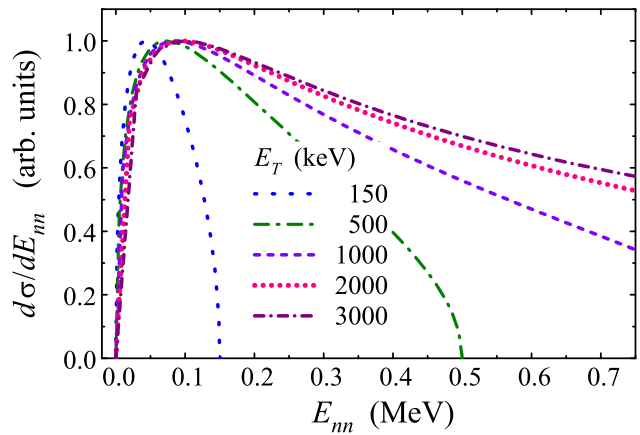}
\end{center}
\caption{The energy distributions for relative energy $E_{nn}$ between two neutrons, are shown for different total decay energies $E_T$. The results are for calculations with dominant $[s^2]_0$ configuration in the nuclear interior, see Fig.\ \ref{fig:sructure-eff} (a,d). All curves are normalized to unity maximum value.}
\label{fig:dis-enn}
\end{figure}

For the $n$-$n$ decay of the $[s^2]$ configuration the D2nM is providing expected results with explicit low-energy peak associated with $n$-$n$ final state interaction. With energy increase this peak becomes sharper and sharper in the $\varepsilon$ variable. However, if we plot the energy correlation in terms of real $n$-$n$ relative energy $E_{nn}$, one can see in Fig.\ \ref{fig:dis-enn} that for decay energies $E_T>150$ keV the $E_{nn}$ peak position drifts slowly to higher energies. The peak position is stabilized at energies $E_{nn} \sim 80-90$ keV for $E_T \sim 500$ keV and depends only very weakly on $E_T$ after that. To get the $E_{nn}$ peak values above 100 keV, the decay energies $E_T$ exceeding 5 MeV are required.

For the $n$-$n$ decay of the $[p^2]$ configuration the energy evolution of the decay patterns is much more complicated. First at about $E_T\sim 500 $ keV a kind of ``antidineutron'' is formed, providing peak at $\varepsilon > 0.5$ values. At around $E_T\sim 1 $ MeV in addition to ``antidineutron'' an expected dineutron low-$\varepsilon$ peak arises. At even higher energies $E_T>2-3$ MeV the dineutron peak becomes the dominant feature of the spectrum, but integral intensities in the dineutron and antidineutron configurations are about equal. This evidently reflects the double-hump internal spatial configuration of the $[p^2]$ structure. So, we can conclude here that for decays with sufficiently high decay energies the $n$-$n$ momentum distributions formed by $n$-$n$ FSI  can be used to extract information on the internal nuclear structure.

For the $n$-$n$ decay of the $[d^2]$ configuration the most odd-looking results are obtained. It is clear that the correlation patterns for decay energies above $E_T\sim 1 $ MeV tend to reflect the triple-hump configuration of the WF in the internal region. However, in contrast to the $[p^2]$ case, no pronounced low-energy $n$-$n$ peak is obtained in the whole considered $E_T$ domain.

We observe that, in contrast to common expectations, if we assume that the decay process is \emph{totally governed} by $n$-$n$ FSI, this does not mean that a simple picture with a single low-energy ``dineutron'' peak is obtained. The important prerequisite for the latter is dineutron emission from a $[s^2]$ configuration.


\subsection{System size effect in D2nM}


As we have mentioned above that an important motivation for $n$-$n$ correlation studies was connected with the idea that the spatial size of the emitting $n$-$n$ configuration may be established. As we have shown in Section \ref{sec:struct} the results for emission from $[p^2]$ and $[d^2]$ configurations contain a lot of information about structure and cannot be the  right tool here. Hence we study this aspect of the model using $[s^2]$ configuration decay, demonstrating an easier way for interpretation of results.

\begin{figure}
\begin{center}
\includegraphics[width=0.35\textwidth]{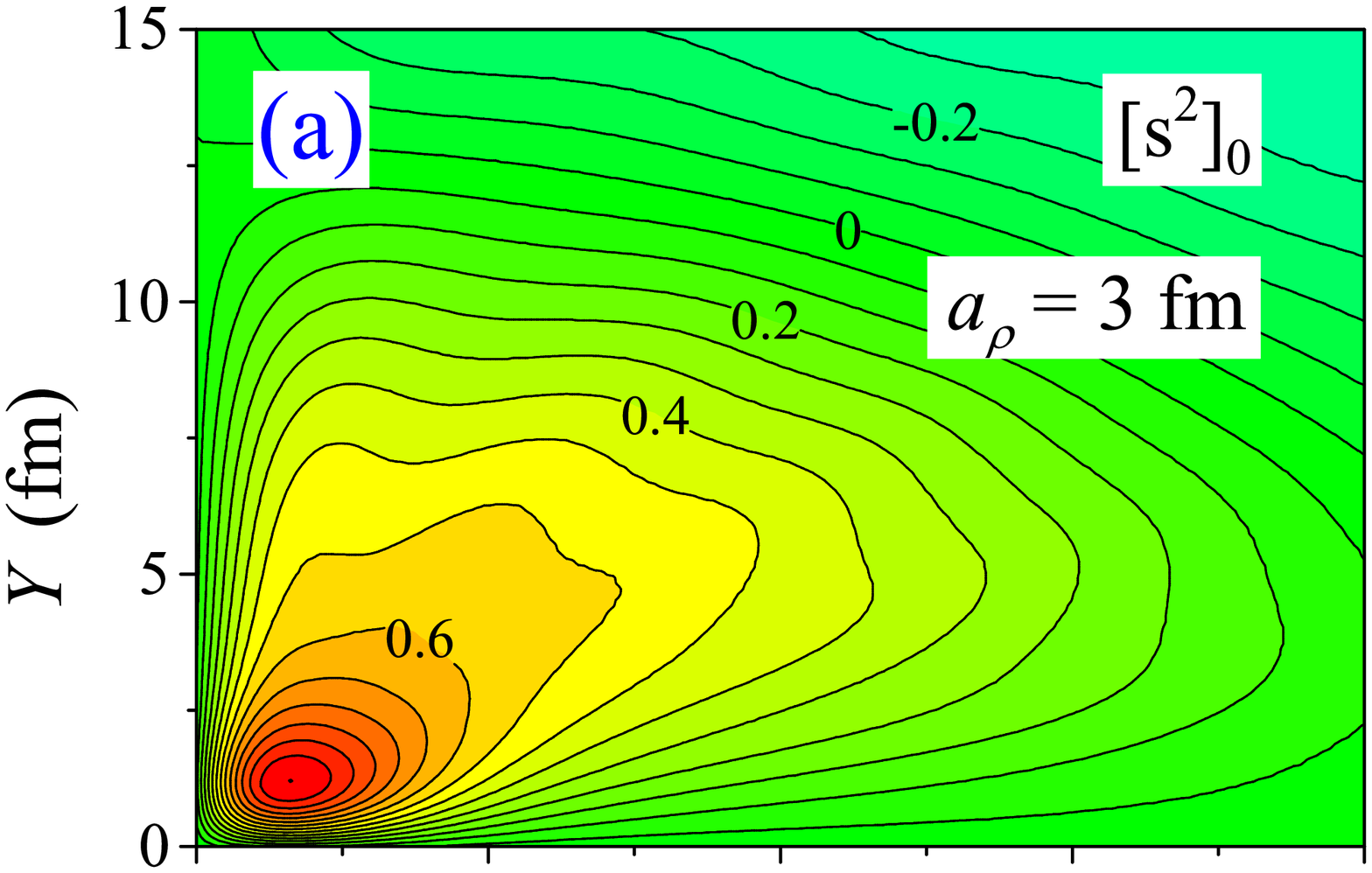}
\includegraphics[width=0.35\textwidth]{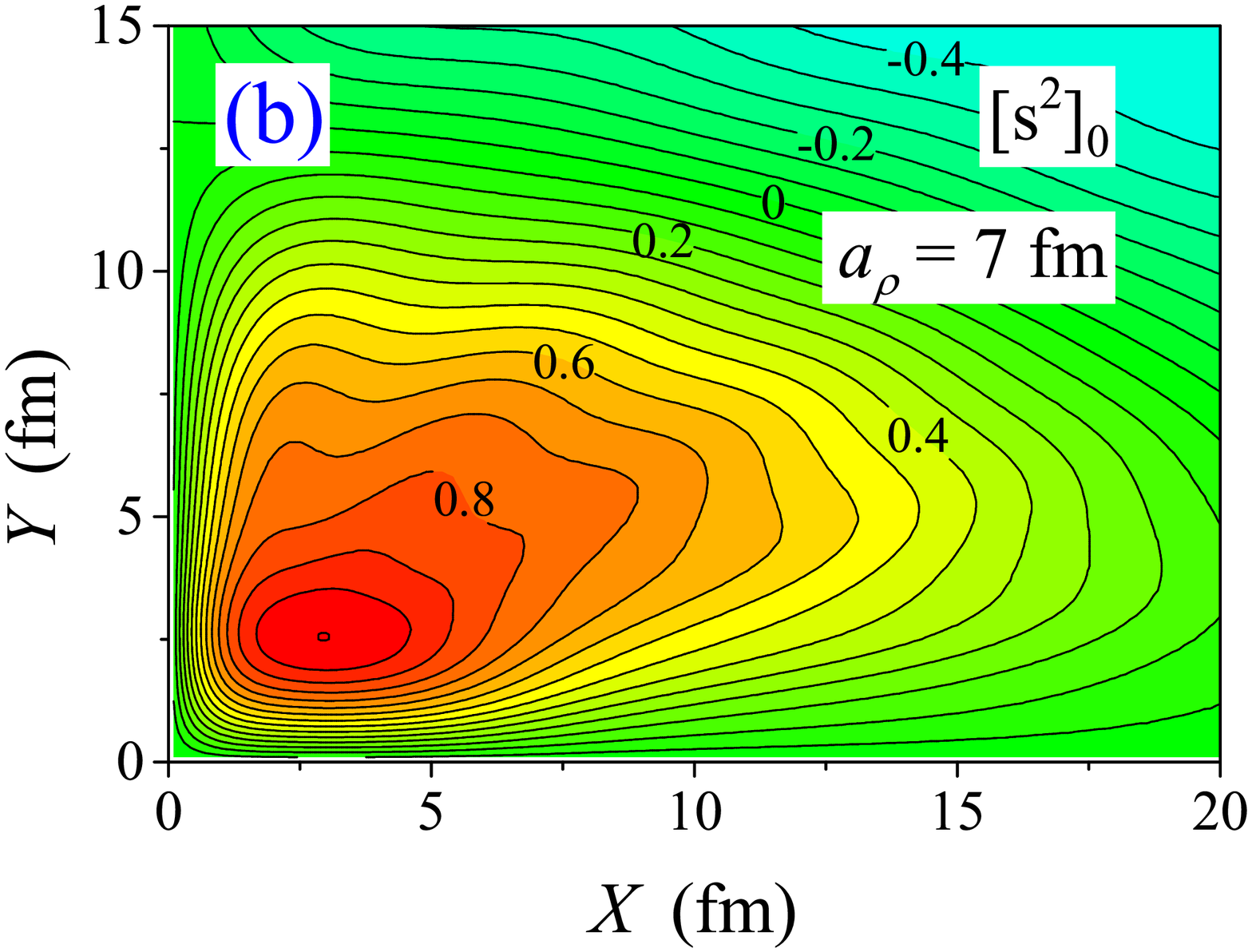}
\end{center}
\caption{Three-body WFs $\Psi^{(+)}$ (real part) calculated with different three-body potential width parameters $a_{\rho}$ (equal 3 and 7 fm) and $E_T=150$ keV. All surfaces are normalized to unity maximum value.}
\label{fig:wfs-dif-rad}
\end{figure}

\begin{figure}
\begin{center}
\includegraphics[width=0.47\textwidth]{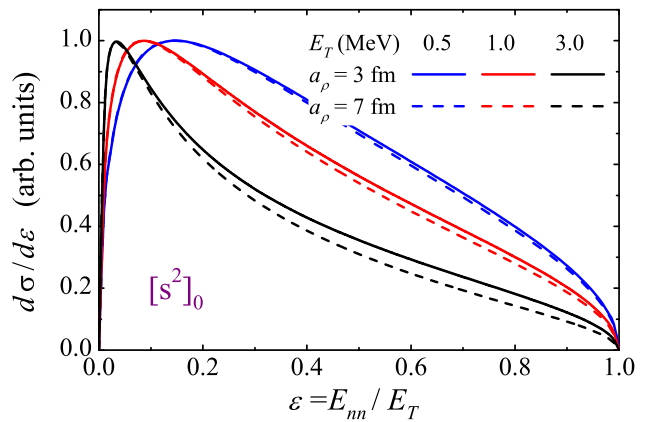}
\end{center}
\caption{Dineutron sensitivity to the size of the emitting 3-body system. The energy distributions between two neutrons are shown for two different three-body potential size parameters $a_{\rho}$ (equal 3 and 7 fm) and for different total decay energies $E_T$. All curves are normalized to unity maximum value.}
\label{fig:dis-eps-rad}
\end{figure}

To vary the nuclear system size we have performed calculations with three-body potential $V_3$ radius chosen to be strongly different from that in Table \ref{tab:potent} Section \ref{sec:struct}. The real part of three-body WFs $\Psi^{(+)}_{E_T}$ obtained with $a_{\rho} = 3$ fm and $a_{\rho} = 7$ fm are shown in Fig.\ \ref{fig:wfs-dif-rad}. It is clear that the radial extent of the nuclear system in the two cases is drastically different. The energy distributions between two neutrons associated with dineutron emission are given in Fig.\ \ref{fig:dis-eps-rad} for three different total decay energies $E_T$. We see that variation of the size of the emitting system practically does not affect the $n$-$n$ correlations.

\subsection{System geometry effect in S2nM}

\begin{figure}
\begin{center}
\includegraphics[width=0.259\textwidth]{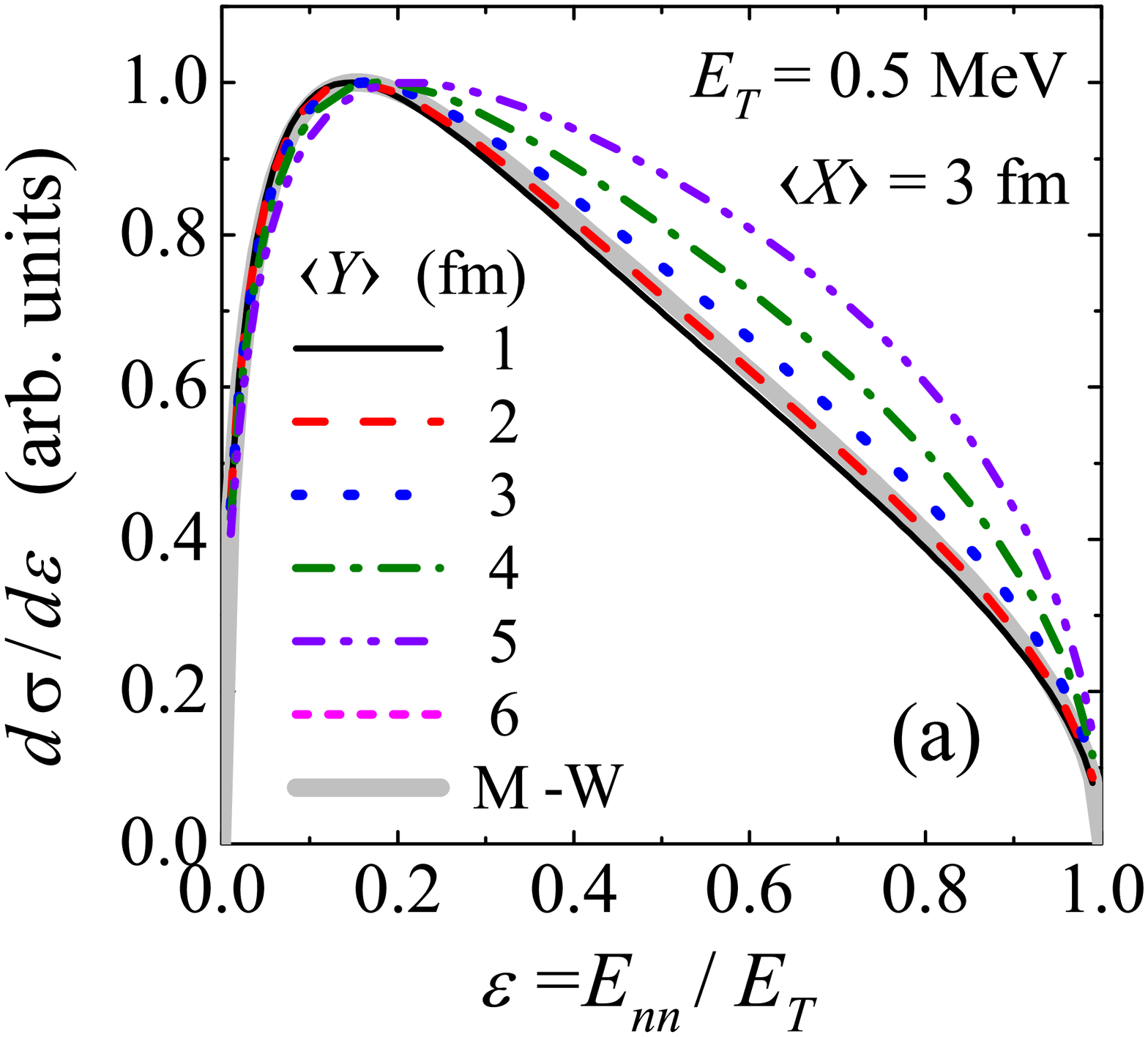}
\includegraphics[width=0.216\textwidth]{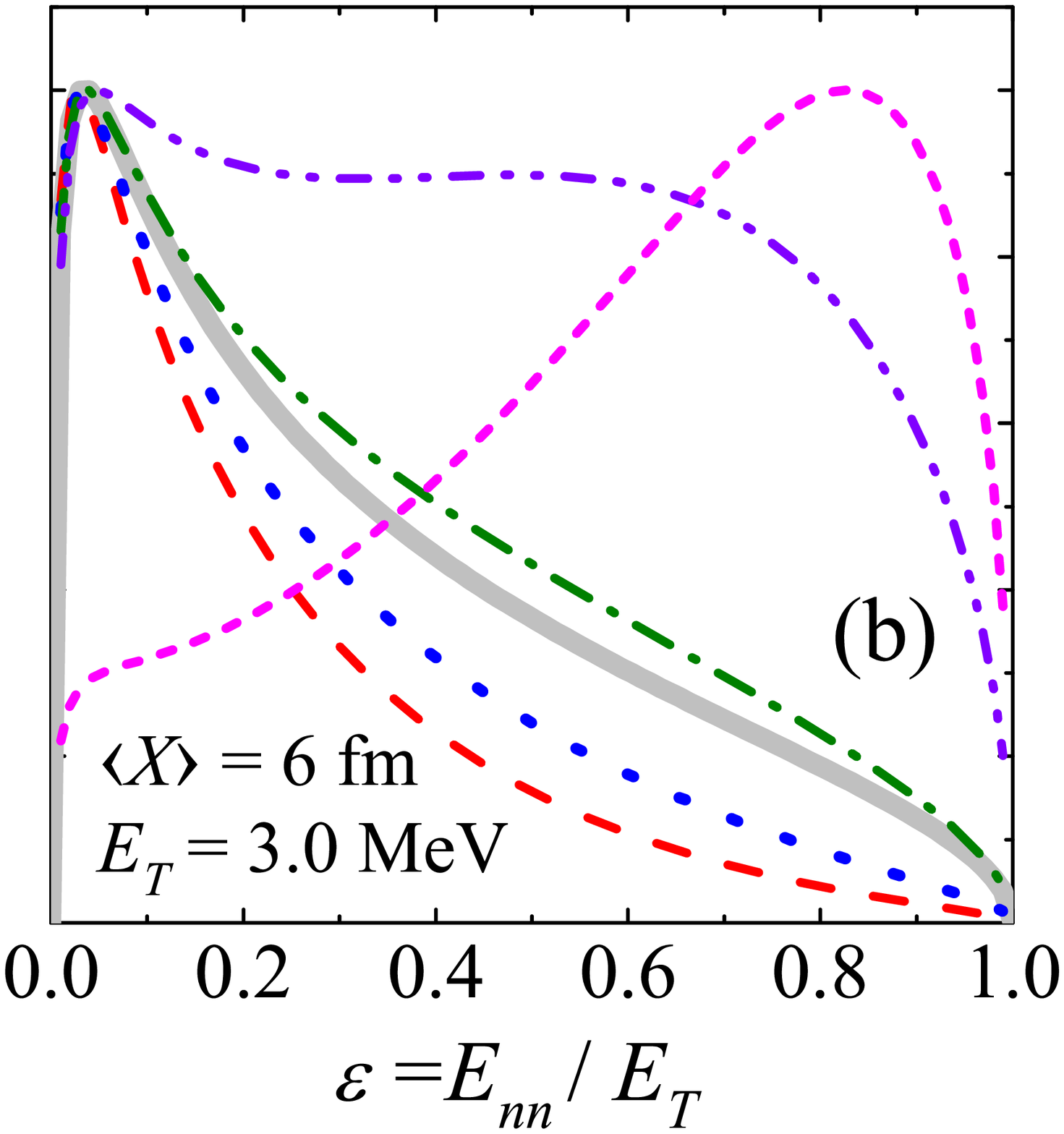}
\end{center}
\caption{Neutron-neutron energy correlation for S2nM emission of dineutron from sources with different three-body T-system ``geometries''. Two cases are illustrated: (a) $E_T=0.5$ MeV, $\langle X \rangle = 3$ fm and (b) $E_T=3$ MeV, $\langle X \rangle = 6$ fm, while the rms $\langle Y \rangle$ distance is varied. All curves are normalized to unity maximum value.}
\label{fig:sdn-vs-mw}
\end{figure}

The observation of the previous Subsection is in strong contrast to expectations. How could it be that the distance between neutrons in the emitting source does not affect the observed $n$-$n$ correlations? It can be understood recalling that in the method used for variation of the nuclear size we actually vary the $\rho$ value for the whole system. This means that we synchronously change both the mean sizes in $X$ and $Y$ coordinates. Let us consider analytic source function (\ref{eq:factor}) for static emission of a dineutron which allows to vary the ratio $\langle X \rangle / \langle Y \rangle$:
\begin{equation}
\Phi(\mathbf{X},\mathbf{Y}) = \Phi(\mathbf{X}) \Phi(\mathbf{Y}) \,,
\label{eq:sour-fact}
\end{equation}
where the radial functions $\Phi(\mathbf{r})$ are defined by Eq.\ (\ref{eq:source-pp}). The results are shown in Fig.\ \ref{fig:sdn-vs-mw} and they really demonstrate that even for emission from pure $[s^2]_0$ configuration, a broad variety of different energy distributions is possible. Here we have to conclude that in contrast with common expectations, even for  emission from $[s^2]$ configuration the dineutron correlation is sensitive not so much to the mean distance $\langle X \rangle$ of the emitting source, but to the ``geometry'' of the source --- the ratio of $\langle X \rangle$ and $\langle Y \rangle$.

\subsection{Static emission of dineutron vs.\ Migdal-Watson approximation}
\label{sec:mw-valid}

\begin{figure}
\begin{center}
\includegraphics[width=0.47\textwidth]{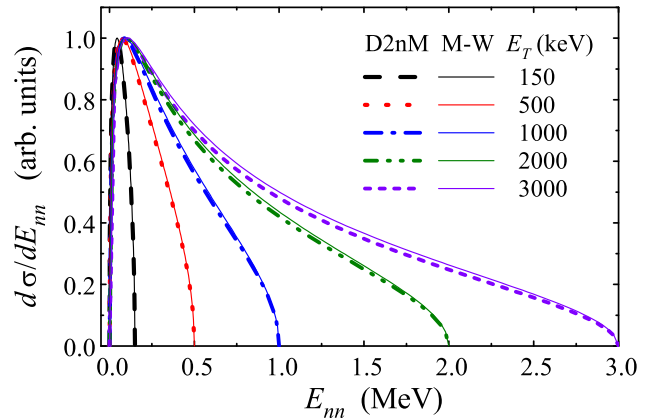}
\end{center}
\caption{Energy distributions for relative energy $E_{nn}$ between two neutrons calculated in D2nM, are given for different total decay energies $E_T$ and compared with Migdal-Watson approximation (thin solid curves of the same color). The calculations correspond to dominant $[s^2]_0$ configuration in the nuclear interior. All curves are normalized to unity maximum value.}
\label{fig:dis-enn-mw}
\end{figure}
\begin{figure}
\begin{center}
\includegraphics[width=0.45\textwidth]{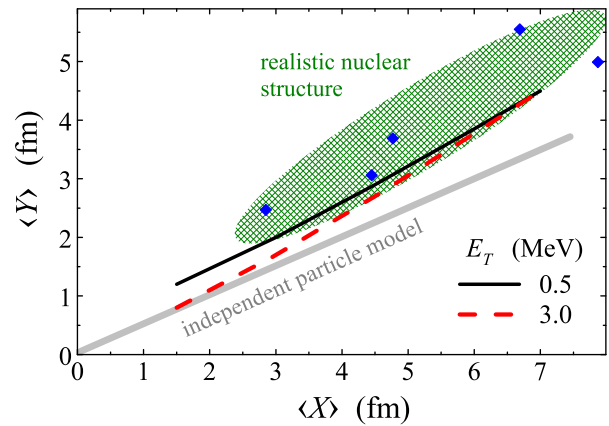}
\end{center}
\caption{Ratio between $\langle X \rangle$ and $\langle Y \rangle$ values at which the $E_{nn}$ energy correlation for S2nM emission from a $[s^2]$ configuration coincides with one obtained in Migdal-Watson approximation. Solid and dotted curves correspond to total decay energies $E_T$ equal 0.5 and 3 MeV respectively. The hatched region qualitatively corresponds to realistic relations between $\langle X \rangle$ and $\langle Y \rangle$ values. The nuclei mentioned in Table \ref{tab:geometries} are shown by blue diamonds.}
\label{fig:sdn-vs-mw-collect}
\end{figure}

The above calculations demonstrate broad variety of dineutron correlation patterns depending on emission conditions. A natural question here is: Why is the Migdal-Watson picture so widespread used as generic understanding of dineutron emission phenomena? Fig.\ \ref{fig:dis-enn-mw} compares the energy distributions obtained in the D2nM for $[s^2]_0$ case with Migdal-Watson results, showing that for different energies they agree extremely well. What is the reason --- is this type of correlation by necessity obtained for emission from a $[s^2]_0$ configuration? If we study systematically the correlation dependence on geometry of the source for static dineutron emission, the reason becomes clear. Fig.\ \ref{fig:sdn-vs-mw} shows examples of correlation evolution for systematic variation of the source geometry in S2nM. There is a broad variety of possible correlation pictures. We find, however, that for certain geometries, namely, for
\begin{equation}
\langle Y \rangle \, \lesssim \,\langle X \rangle \,\lesssim \, 2\langle Y \rangle \,,
\label{eq:geom-ratio}
\end{equation}
the correlations vary quite slowly and approach the Migdal-Watson results. It is clear that we can define a ratio of $\langle X \rangle$ and $\langle Y \rangle$ values such that S2nM results coincide with Migdal-Watson, see Fig.\ \ref{fig:sdn-vs-mw-collect}.

To interpret these results we should recall that for an independent particle model with two nucleons populating the same orbital configurations the condition $\langle X \rangle = 2 \langle Y \rangle$ is satisfied (or what is the same, the average angle between two nucleons is equal to $\pi/2$). In reality the nucleon-nucleon interaction leads to formation of a closer configuration of two nucleons in the nuclear interior, which is also often referred to as ``dineutron''. This leads to smaller $\langle X \rangle/\langle Y \rangle$ values compared to that in the independent particle model. The realistic values reside exactly in the range given above by Eq.\ (\ref{eq:geom-ratio}). To illustrate this statement the calculated geometrical characteristics of some two-nucleon halo systems are provided in Table \ref{tab:geometries}. The geometry of continuum WFs obtained in the D2nM can be roughly estimated via the WF main peak position in the $\{X,Y\}$ plane, see Figs.\ \ref{fig:sructure-eff} (a) and \ref{fig:wfs-dif-rad}. It also satisfies the condition in Eq.\ (\ref{eq:geom-ratio}).

We conclude that the Migdal-Watson approximation for dineutron emission ($n$-$n$ FSI totally defines the decay dynamics) works nearly perfect when the $[s^2]_0$ source geometry is defined by formation of spatial ``dineutron'' correlation induced by pairing interactions in the internal nuclear region.

\begin{table}[b]
\caption{T-geometry of several bound three-cluster (e.g.\ two-nucleon halo) systems residing near the dripline obtained in the three-body cluster model calculations. The $^{3}$H properties are trivially inferred from experimental data on charge radius.}
\begin{ruledtabular}
\begin{tabular}[c]{cccccc}
Nucleus & Model & $\langle X \rangle$  & $\langle Y \rangle$ &
$\langle X \rangle / \langle Y \rangle$ & Ref. \\
\hline
$^{3}$H   & $n$+$n$+$p$      & 2.85 &  2.47 & 1.15 & \cite{Angeli:2013} \\
$^{6}$He  & $^4$He+$n$+$n$   & 4.77 &  3.69 & 1.29 & \cite{Zhukov:1993} \\
$^{11}$Li & $^9$Li+$n$+$n$   & 6.69 &  5.55 & 1.21 & \cite{Shulgina:2009} \\
$^{17}$Ne & $^{15}$O+$p$+$p$ & 4.45 & 3.06  & 1.45 & \cite{Grigorenko:2003}  \\
%
%
$^{22}$C  & $^{20}$C+$n$+$n$ & 7.87 & 4.99  & 1.58 & \cite{Ershov:2012}  \\
\end{tabular}
\end{ruledtabular}
\label{tab:geometries}
\end{table}


\section{Discussion}


\subsection{General}


The emission of two nucleons is often discussed in terms of a dominating ``diproton'' or ``dineutron'' decay mechanism. In this work we have tried to bring some clarity to the issue by constructing a model which allows to explicitly isolate effect of the nucleon-nucleon final state interaction. Based on the obtained results we can conclude that from a theoretical formal point of view the common vision of ``dineutron'' as a low-energy enhancement in the the nucleon-nucleon energy distribution is not substantiated.

It seems that in the discussions of dinucleon emission there is some misunderstanding about relation of \emph{necessary} and \emph{sufficient} conditions. If we \emph{observe} low-energy emission enhancement in the nucleon-nucleon spectrum this enhancement is evidently connected with $N$-$N$ FSI. This condition can be regarded as necessary, because the huge scattering length in the $N$-$N$ channel ($\sim 20$ fm) means that in nuclear physics we do not have systems which can emit nucleons in such a way that they are outside the FSI range. This thing is unavoidable and thus trivial. However, as we have shown in this work for various emission conditions, the presence of $N$-$N$ FSI as the only factor governing two-nucleon emission does not lead to a unique result (low-energy emission enhancement in the nucleon-nucleon spectrum). Even in the simplified dineutron theoretical model the major factors defining the nucleon-nucleon relative energy distributions in the final state are structure and spatial distributions in the internal region.

This result strongly discourages discussion of nucleon-nucleon correlations, observed in reactions and decays, in loosely defined terms such as a ``diproton'' or ``dineutron'' reaction mechanisms. In contrast it supports our confidence that comprehensive treatment of three-body decay mechanisms in all their complexity is a promising approach for extraction of information about nuclear interior and reaction mechanisms.


\subsection{Lifetimes in the D2nM by example of $^{26}$O}

\begin{figure}
\begin{center}
\includegraphics[width=0.47\textwidth]{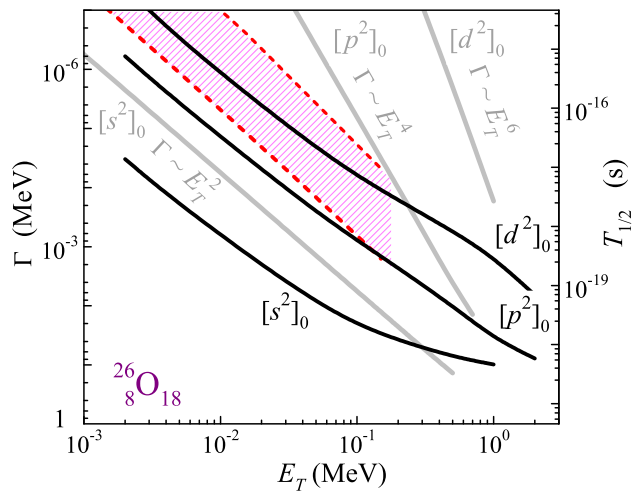}
\end{center}
\caption{Lifetime of the $^{26}$O g.s.\ in D2nM (black curves) for different structure assumptions are compared with direct decay model estimates of \cite{Grigorenko:2011} (gray curves) and three-body calculations \cite{Grigorenko:2013} (hatched area between red dashed curves).}
\label{fig:lifetime}
\end{figure}

Here we consider how the lifetimes obtained in D2nM are compared with results of different decay models. This is illustrated by example of $^{26}$O g.s.\ $2n$ decay, see Fig.\ \ref{fig:lifetime}.

The ``direct decay model'' estimates \cite{Grigorenko:2011} assume independent emission of nucleons from definite shell configurations. This model contains sensitivity to interactions in the core-nucleon channel, while the nucleon-nucleon FSI is neglected. The D2nM results provide similar dependence of the decay width on energy in a broad energy range in the assumption about direct emission of nucleons off $[s^2]_0$ configuration. However, the decay is about one order of the magnitude faster in the case of D2nM. This is evidently connected to additional boost for $2n$ penetration due to $n$-$n$ interaction in the subbarrier region. This observation is also consistent with results of $2p$ decay studies: the ``diproton decay'' estimates are providing the largest width values among all models, typically considerably overestimating widths relative to experiment \cite{Pfutzner:2012}.

Three-body model calculations of $^{26}$O decay from Ref.\ \cite{Grigorenko:2013} demonstrated strong sensitivity of width to details of core-$n$ interactions, indicated by hatched area between red dotted curves in  Fig.\ \ref{fig:lifetime}. It can be seen in the Figure that D2nM calculations with realistic assumption about $[d^2]$ structure of $^{26}$O g.s.\ provide results consistent with complete three-body model calculations. So, application of the D2nM for lifetime estimates seems to be correct within an order of magnitude.

\subsection{Correlations in the decay of $^{26}$O}

\begin{figure}
\begin{center}
\includegraphics[width=0.235\textwidth]{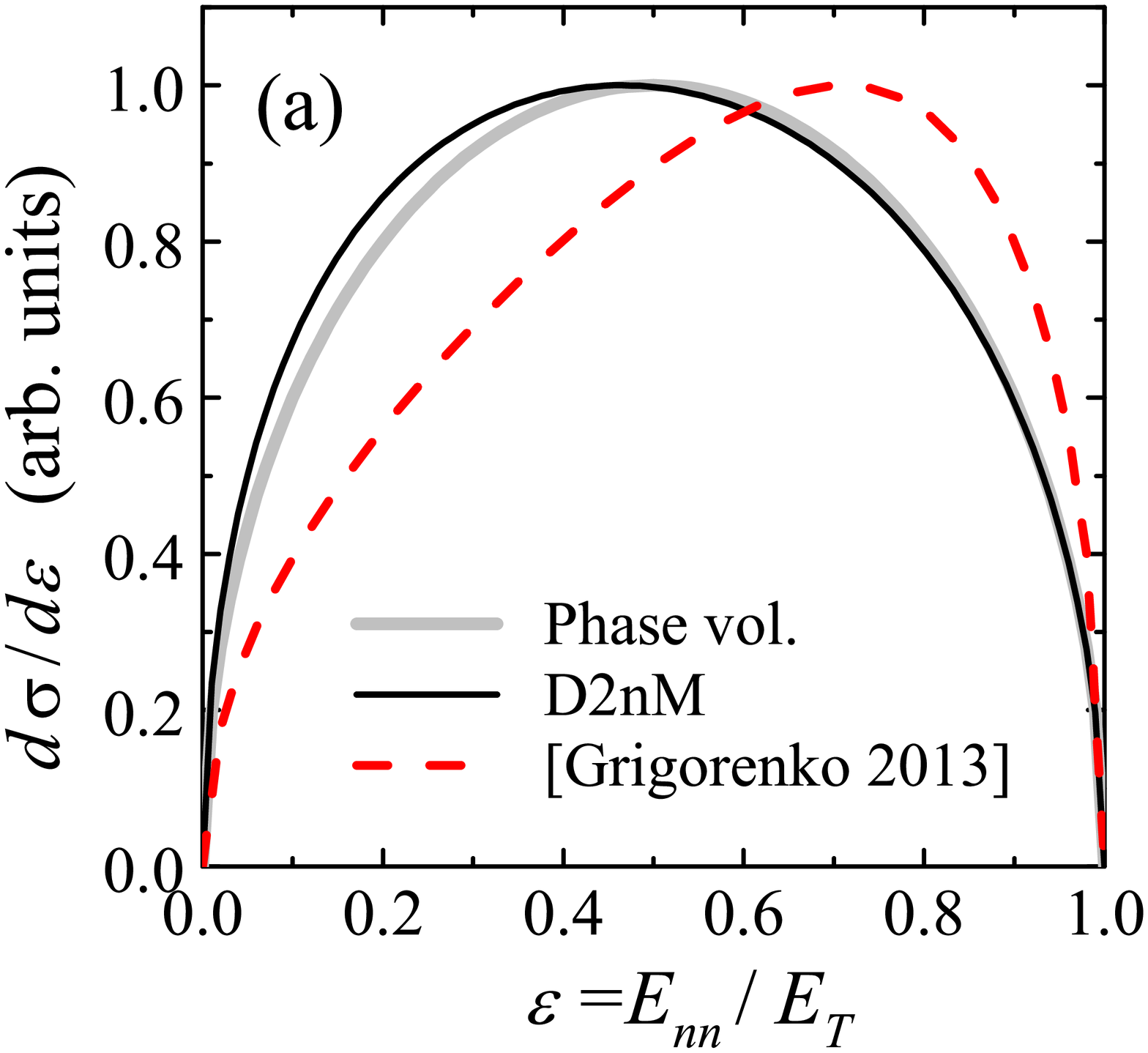}
\includegraphics[width=0.235\textwidth]{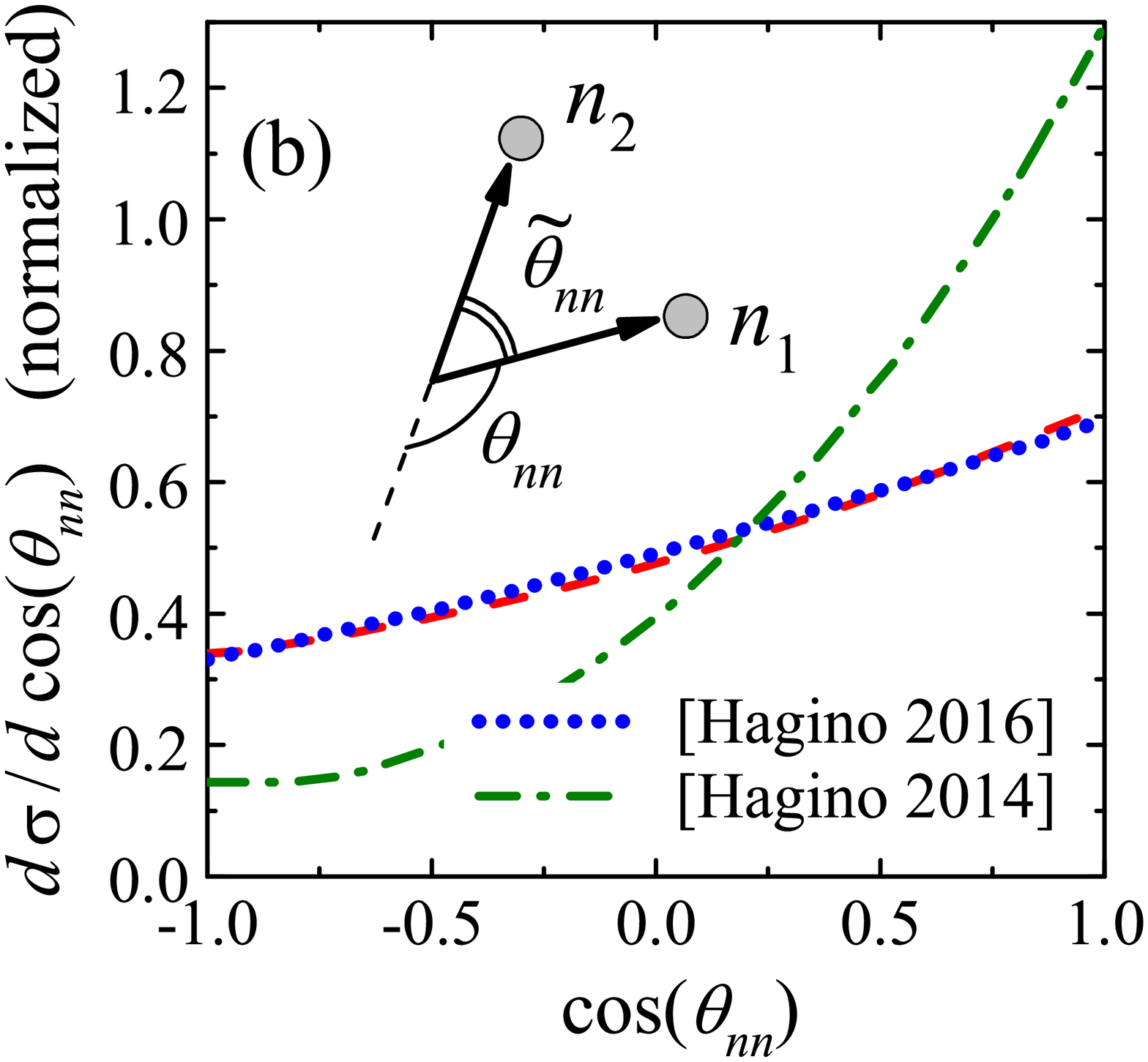}
\end{center}
\caption{Neutron-neutron correlations in the decay of $^{26}$O g.s.\ in D2nM, $E_T=150$ keV --- black curve in panel (a), in three-body models of Refs.\ \cite{Grigorenko:2013}, $E_T=75$ keV --- red dashed curves in panels (a) and (b), \cite{Hagino:2014}, $E_T=150$ keV --- green dash-dotted curve in panel (b), and \cite{Hagino:2016}, $E_T=150$ keV --- blue dotted curve in panel (b). Energy distributions are normalized to unity maximum value; angular distributions are normalized for integration over $d \cos (\theta_{nn})$.}
\label{fig:corel-o26}
\end{figure}

Another issue for D2nM is how correlations between neutrons compare to results obtained in different models. This is illustrated by example of low-energy  $2n$ decay of the $^{26}$O ground state, see Fig.\ \ref{fig:corel-o26}. This figure shows both the energy correlations for parameter $\varepsilon=E_{nn}/E_T$ and angular correlations for angle $\theta_{nn}$. The hyperspherical method provides convenient instruments for construction of all possible types of correlations
\cite{Pfutzner:2012}. The $\varepsilon$ and $\theta_{nn}$ correlations are not independent and reflect the same type of correlation dynamics in different representations. For consistency with our previous works (e.g.\ Refs.\ \cite{Pfutzner:2012,Grigorenko:2013}) $\theta_{nn}$ is defined as the angle between momenta $\mathbf{k}_{n_1}$ and $-\mathbf{k}_{n_2}$. It should be noted that angle $\theta_{nn}$ in Fig.\ \ref{fig:corel-o26} is defined as $\pi-\tilde{\theta}_{nn}$, where $\tilde{\theta}_{nn}$ is angle in \cite{Hagino:2014,Hagino:2016}.  It can be seen that all the previous three-body model calculations \cite{Grigorenko:2013,Hagino:2014,Hagino:2016} predict similar correlations behavior which can be interpreted as effective repulsion between neutrons in the final state (average angle between neutron emission directions is more than 90 degrees). In contrast the D2nM predicts small effective attraction: The peak in the energy distribution is shifted to slightly smaller $\varepsilon$ values than for the ``phase space'' distribution shown for reference in Fig.\ \ref{fig:corel-o26} (a). So, for \emph{correlations}, the ``dineutron'' assumption provides a qualitatively wrong trend in the case of low-energy $^{26}$O g.s.\ decay.

\subsection{Correlations in the decay of $^{5}$H}

\begin{figure}
\begin{center}
\includegraphics[width=0.47\textwidth]{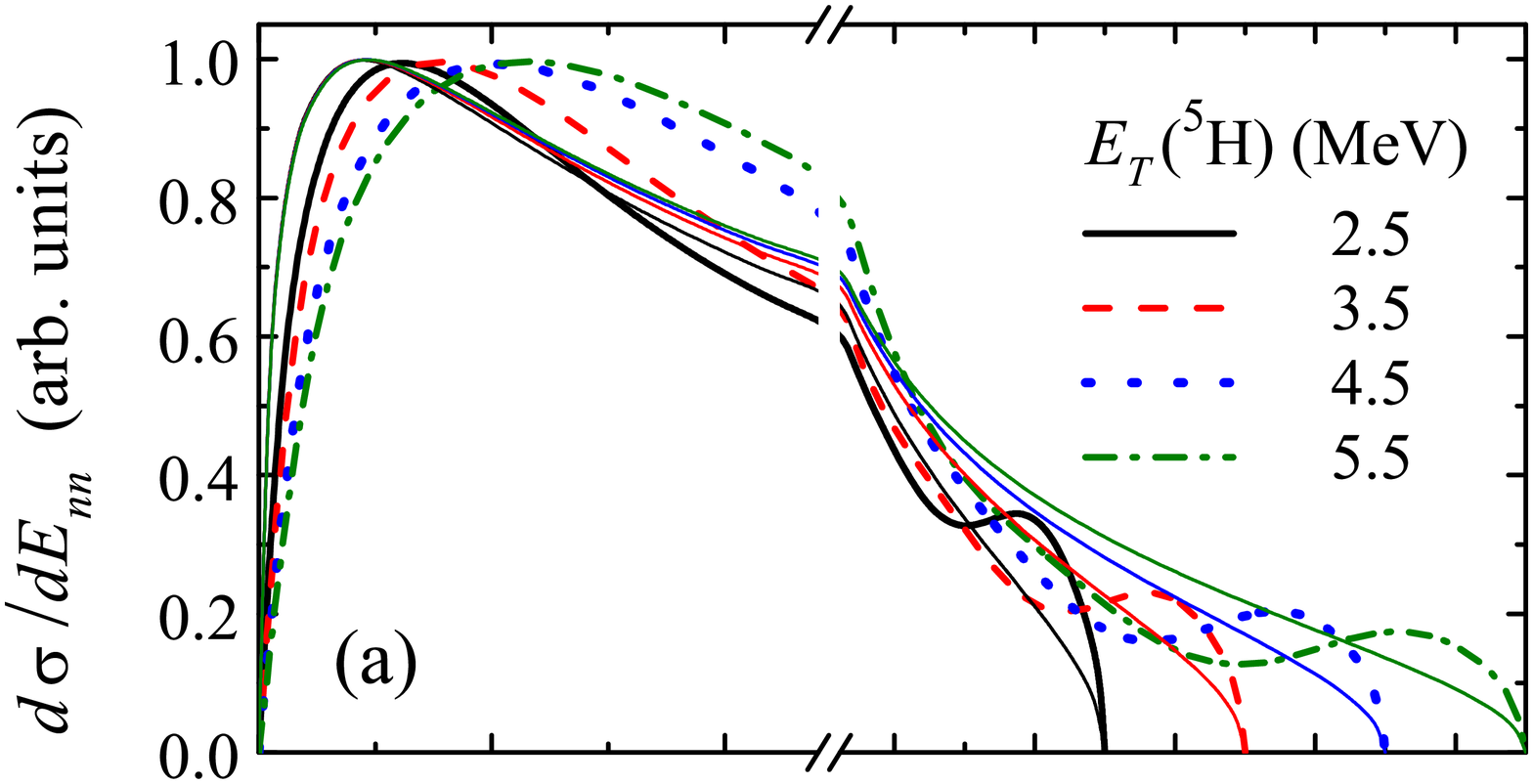}
\includegraphics[width=0.47\textwidth]{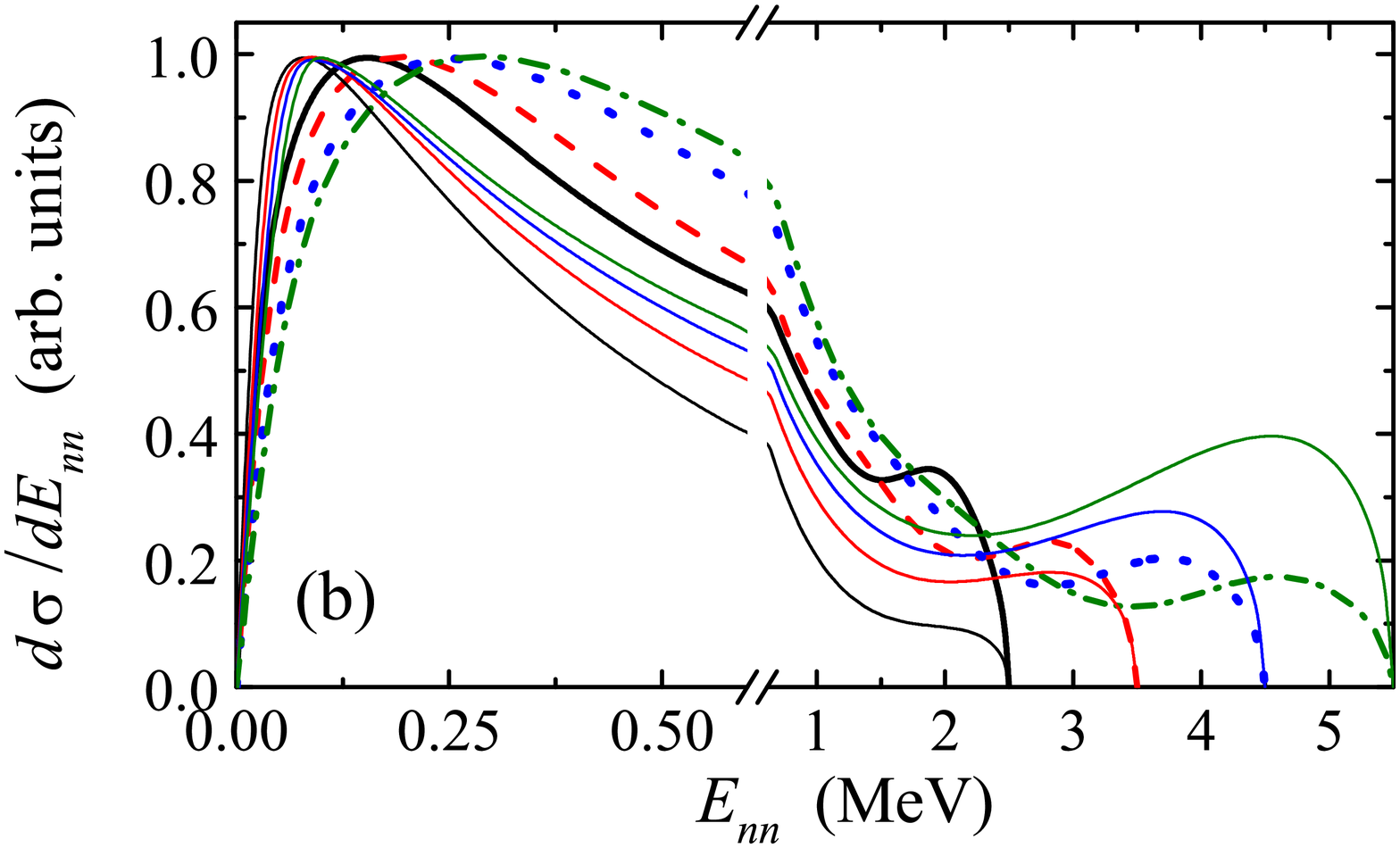}
\end{center}
\caption{The experimental $n$-$n$ relative energy spectra for $^{5}$H from  Ref.\ \cite{Golovkov:2005} reconstructed for different total $^{5}$H decay energies $E_T$. Thin solid lines of the same color in panel (a) show the Migdal-Watson approximation for the same energy. The D2nM calculations for decay of $[p^2]_2$ configuration are given in panel (b). All curves are normalized to unity maximum value.}
\label{fig:corel-5h}
\end{figure}

D2nM calculations for decay of a $[s^2]$ configuration demonstrate nice agreement for $n$-$n$ with Migdal-Watson approximation, see Fig.\ \ref{fig:dis-enn-mw}.
Also we found that for low total decay energies $E_T>150$ keV some kind of scaling behavior is obtained, see Fig.\ \ref{fig:dis-enn}: The peak in the $E_{nn}$ spectrum slowly drifts to higher energies with total decay energy $E_T$ increase. It is interesting to note that analogous scaling behavior was observed in the studies of two-neutron decay of $^{5}$H \cite{Golovkov:2005}, see Fig.\ \ref{fig:corel-5h}. The $E_{nn}$ relative energy spectra were carefully reconstructed in this work for several decay energies of $^{5}$H.  However, what we see is that the drift of the $E_{nn}$ peak to higher energy in data continues up to $E_T=5$ MeV --- the maximal energy obtained in this experiment. So, we see that $E_T$ evolution of low-energy peak in Migdal-Watson approximation is strongly different from the experimentally observed picture.

Some precaution is needed here because the majority of the mentioned spectrum is connected with the decay of excited states of $^{5}$H expected to have $[p^2]_2$ orbital configuration. Within D2nM it is possible to study the decay specifically of this configuration. Compared to Migdal-Watson results (and data as well) the D2nM provides here the low-energy peak even at lower energies. The double-hump structure connected with decay of $[p^2]_2$ configuration and observed in the decay of $^{5}$H is present in D2nM results. However, the calculated energy trend predicts enhancement of the large $\varepsilon$  hump with energy $E_T$, while in experiment decrease was actually observed.

Thus none of the predicted ``dineutron'' trends is supported by the experimental data.

\subsection{HBT-like approaches to $n$-$n$ correlations}


The neutron-neutron correlations in the ``trivial dineutron'' treatment of Section \ref{sec:2n-triv} depend only on one parameter --- the radial size of the neutron source. If we integrate the S2nM correlation spectrum (see Section \ref{sec:2n-stat}) over the momentum connected with motion in the $Y$ variable, we retain only the information about neutron-neutron relative distance [this is especially evident for the factorized source (\ref{eq:factor})]. This fact defines the use of intensity interferometry approach as a femtoscopy tool in high-energy physics. The idea to use the $n$-$n$ correlations in the decays following reactions with exotic nuclei has two obstacles.

\noindent (i) For decay of higher shell configurations (such as $[p^2]$ and $[d^2]$) even such an integral correlation information cannot be straightforwardly related to radial characteristics of the source.

\noindent (ii) Technically the existing experimental setups are arranged in such a way that their acceptance for $2n$ events is drastically falling with total energy of two neutrons in the projectile frame. This fall is typically taking place in the energy range 1--3 MeV. For this reason we consider the two-neutron events with energy maximum $E_T$ value of 3 MeV in this work. We have demonstrated that $n$-$n$ correlations with such fixed total decay energy could be sensitive to structure, geometry, but not to radial size of the spatial $n$-$n$ correlation.

These issues probably makes the interpretation of the neutron-neutron correlation data in Refs.\ \cite{Marques:2000,Marques:2001,Marques:2002} not quite consistent.


\section{Conclusions}


The dineutron emission is studied in this work in three different models, each with application to a certain realistic scenario. The new development we introduce in this work is the Dynamic Dineutron Model (D2nM) which combines semirealistic internal structure for the nuclear interior with a nucleon-nucleon interaction solely governing the emission process. This model is a subset of the complete three-body problem which allows nice illustration of an efficiently isolated ``dineutron emission'' aspect of this problem. We argue that if we discuss the dineutron emission at all, this should be within a formally correct realization of a theoretical description for such a process.

The results of this work require to critically reconsider several issues which are essential for current investigations. In particular we have demonstrated the following:

\noindent (i) The low-energy $n$-$n$ correlation is typically a testing ground for indications of ``dineutron emission''. We have to state that from a formal point of view a broad variety of ``dineutron'' correlation patterns is possible. A single low-energy peak, even within the simplified D2nM assumptions, should be the indication of emission from $[s^2]$ configuration strictly \emph{with certain geometry}.

\noindent (ii) We found that the idea to define the size of the emitting region via $n$-$n$ correlations, inherited from the HBT-like approach in high-energy physics, does not work for nuclear decays and reactions where sources have definite shell structure and spin-parity. Even for emission from $[s^2]$ configuration, dineutron correlation is sensitive to the exact geometry of the internal WF (e.g. average angle between neutrons).

\noindent (iii) We have given an illustrative explanation why the Migdal-Watson approach works well in the nuclear systems. It is shown that Migdal-Watson-like correlation patterns originate from $[s^2]$ configurations, where two neutrons are more focused in space than in the independent particle case. This is a natural WF geometry effect of the attractive pairing interaction in a nucleus and thus such a spatial configuration is typically ``pre-conditioned'' for many processes of two-neutron emission.

%
\begin{acknowledgments}
%

L.V.G.\ was partly supported by the Russian Science Foundation (grant
No. 17-12-01367). We are grateful Prof.\ I.G.\ Mukha for useful comments.

\end{acknowledgments}


\bibliographystyle{apsrev4-1}
\bibliography{d:/latex/all}


\end{document}